\documentstyle[preprint,prl,aps,epsf]{revtex}

\begin{document}
\def\lsim{\mathrel{\raise.3ex\hbox{$<$\kern-.75em\lower1ex\hbox{$\sim$}}}}
\def\gsim{\mathrel{\raise.3ex\hbox{$>$\kern-.75em\lower1ex\hbox{$\sim$}}}}
\def\fbi{~{\rm fb}^{-1}}
\def\fb{~{\rm fb}}
\def\pbi{~{\rm pb}^{-1}}
\def\pb{~{\rm pb}}
\def\mev{~{\rm MeV}}
\def\gev{~{\rm GeV}}
\def\tev{~{\rm TeV}}
\tightenlines
\preprint{
$\vcenter{
\hbox{\bf UCD-2000-06} 
\hbox{\bf LBNL-45042}
\hbox{\bf hep-ph/0002046}
\hbox{January, 2000}
}$
}
\title{Higher Twist Contributions To $R$-Hadron Phenomenology In The Light
Gluino Scenario}
\author{T.D. Gutierrez$^a$, R. Vogt$^{a,b}$\footnote{This work was supported 
in part by
the Director, Office of Energy Research, Division of Nuclear Physics
of the Office of High Energy and Nuclear Physics of the U. S.
Department of Energy under Contract Number DE-AC03-76SF00098.}, 
J.F. Gunion$^a$\footnote{Work supported in part by the U.S. 
Department of Energy and the U.C. Davis Institute for High Energy Physics.}}
\address{$^a$Physics Department, University of California, Davis, CA, 95616}
\address{$^b$Nuclear Science Division, Lawrence Berkeley Laboratory, 
Berkeley, CA, 94720} 
\maketitle
\begin{abstract}
The open light gluino window allows non-trivial higher twist gluino
contributions to the proton wave function.  Using a two-component model
originally developed for charm hadroproduction,
higher twist intrinsic gluino contributions to final state
$R$-hadron formation are shown to enhance leading
twist production in the forward $x_{F}$ region. We calculate $R$-hadron
production at $p_{\rm{lab}}=800$ GeV in $pp$, 
$p$Be, and $p$Cu interactions with light gluino masses of $1.2$, $1.5$, $3.5$, 
and $5.0$ GeV.
\end{abstract}
\vskip .2in
\centerline{PACS Numbers: 12.60.Jv, 13.85.-t, 14.80.Ly}

\section*{Introduction}
The gluino is the supersymmetric partner of the gluon.  It is
an electromagnetically neutral, adjoint fermion with the same color 
structure as its boson
counterpart.  As yet, no clear experimental evidence of supersymmetric
particles has been found.  The most likely reason for this is the large
expected mass of the supersymmetric particles 
($\Lambda_{\rm{SUSY}}\sim 1$ TeV).  
However, an intriguing scenario exists whereby the
gluino is not only the lightest supersymmetric particle but also very light 
compared to the SUSY scale, $m_{\tilde{g}} \ll 100$ GeV. This possibility
arises naturally in a number of quite attractive 
models characterized by special boundary
conditions at the grand unification scale 
\cite{glennys,guniondrees2} and in certain
models of  gauge-mediated supersymmetry breaking \cite{raby}. 

Light gluinos are
predicted to form relatively light bound states of quarks or gluons and
gluinos called $R$-hadrons \cite{Farrar}.  The lightest predicted $R$-hadrons
include mesinos ($q\bar{q}\tilde{g}$), two barinos, $R^{+}$($uud\tilde{g}$) and
$S^{0}$($uds\tilde{g}$), gluinoballs
($\tilde{g}\tilde{g}$), and the glueballino
or $R^{0}$ ($\tilde{g}g$).  The properties of $R$-hadrons including 
their mass, decay modes,
and lifetimes depend strongly on the mass of the gluino.  

There have been many theoretical and experimental attempts to
find evidence for and/or exclude the light gluino scenario. Searches
for $R$-hadrons produced in fixed target experiments
have been performed for a number of the predicted $R$-hadron decay channels 
\cite{KTeV1,KTeV2,E761,CUSB}. Effects of a light gluino on QCD observables have
been analyzed \cite{nogluinos}. Stable particle searches, $\Upsilon$ decays,
beam dump experiments {\it etc.} all have potential sensitivity to the presence
of a light gluino or the $R$-hadrons. A brief summary of the
various possible resulting constraints on a light gluino
is given in Ref.~\cite{Clav}.  In addition,
Ref.~\cite{bcg} claims that $m_{\tilde g}> 2.5-3\gev$
is excluded on the basis of their analysis of OPAL data.
Although these various analyses are, in combination, potentially
sensitive to most regions of light gluino mass, all rely on model-dependent
inputs. As a result, we believe that at present 
it is impossible to {\it definitively} exclude any
gluino mass below  $4-5\gev$. Thus, it is of great interest to find
additional approaches for discovering and/or constraining
light gluinos and the $R$-hadrons.

In this paper, we will explore the possibility of detecting $R$-hadrons
at large $x_F$ in $pp$ and $pA$ fixed-target interactions. Our calculations
will be restricted to the 
$m_{\tilde g}\sim 1.2-5$ GeV region where we can be confident that
the semi-perturbative techniques that we employ are reliable.
This region is of particular phenomenological interest
because of the analogy that can be drawn between heavy quark and light gluino
production.  Indeed, if the gluino and heavy
quark masses are comparable, one might anticipate observation of 
hard gluino production analogous to that 
already observed in high-$x_F$ charm hadroproduction
\cite{leadingcharm}. The leading-twist pQCD predictions for
charm production in $pp$ and $pA$ collisions fail to account
for many features of the high-$x_F$ data.  
These include unexpectedly large production rates and 
anomalies such as flavor correlations between the produced hadrons and the
valence spectators, manifested as leading charm and a strong 
$D^{+}/D^{-}$ asymmetry in $\pi^- A$ interactions \cite{leadingcharm}, 
double $J/\Psi$ production at large $x_{F}$ \cite{NA3psipsi}, and
Feynman scaling of $J/\Psi$ production in $pA$ interactions
\cite{NA3_1,psiadep}, 
all of which suggest 
a breakdown of factorization \cite{fact} at large $x_{F}$.   
The anomalies and cross section enhancement
may be  partly explained by higher twist terms in the operator
product expansion (OPE) on the light cone associated with the dynamics of the 
QCD bound state. Analgous terms should be present for light gluinos.

The intrinsic charm model (IC) \cite{StanIC,BGSIC} 
approximates non-perturbative higher twist
Fock-state contributions of heavy quarks in hadronic wave functions.  The
phenomenological predictions of IC directly address the above puzzles in
charm hadroproduction \cite{BVIC1,BVIC2,BVIC3,BVIC4,BVIC5,GutV}.
For example, IC provides a coalescence mechanism whereby final state hadrons 
can share valence quarks with the projectile, naturally producing leading 
particles.

In analogy with leading charm, we study $R$-hadron distributions using 
``intrinsic
gluinos'' (I$\tilde{\rm{G}}$) in regions of phase space where the gluino 
mass and momentum fractions conspire so that higher twist effects cannot
be ignored.  In this paper, we calculate enhancements over the
leading twist $R$-hadrons $x_{F}$ distributions with 
gluino masses $m_{\tilde{g}}=1.2$, $1.5$, $3.5$, and $5.0$ GeV.
Both $pp$ and $pA$ interactions at $p_{\rm{lab}}=800$ GeV are considered.

\section*{pQCD Light Gluino Hadroproduction}

In pQCD, gluinos are produced in pairs by $gg$ fusion and $q\bar{q}$
annihilation, $gg,q\bar{q}\rightarrow\tilde{g}\tilde{g}$, 
as well as quark-gluon
scattering to squark and gluino, $qg\rightarrow\tilde{q}\tilde{g}$.   
Precision Z-pole data has
constrained the squark mass to be greater than $100$ GeV, quite large compared
to the light gluino masses considered here.  Therefore, we expect that the $qg$
contribution with the virtual squark in the $t$-channel 
will be small compared to
the other contributions, particularly at fixed-target energies.

The leading twist inclusive $R$-hadron $x_{F}$ distribution at leading order
is obtained from the gluino
$x_{F}$ distribution ($x_{F}=(2 m_{T}/\sqrt{s})\sinh{y}$) which
has the factorized form in pQCD
\begin{equation}
\frac{d\sigma}{dx_{F}}=\sum_{i,j}\frac{\sqrt{s}}{2}\int dz_{3} dy_{2}
d^{2}p_{T} \frac{1}{E_{1}}
\frac{D_{H/\tilde{g}}(z_{3})}{z_{3}} f^{A}_{i}(x_{a})f^{B}_{j}(x_{b})
{1\over\pi}\frac{d\hat{\sigma}_{ij}}{d\hat{t}} \, \, .
\label{dsdxflt}
\end{equation}
Here $a$ and $b$ are the initial partons from projectile and target hadrons $A$
and $B$, $1$ and $2$ are the produced gluinos, and $3$ is the final-state
$R$-hadron.  The sum over $i$ and $j$ extends over all partonic gluino
production subprocesses.  A $K$ factor of 2.5 is included to account 
for NLO corrections.  Since the $K$ factor is approximately constant with 
$x_{F}$ for charm production except as $x_F \rightarrow 1$, 
we assume that the $K$ factor for gluino production is also
independent of $x_F$. 

The fragmentation
functions, $D_{H/\tilde{g}}(z)$ with $z=x_{H}/x_{\tilde{g}}$, describe the
collinear fragmentation of final state $R$-hadrons from the produced gluinos.  
For simplicity, a delta function was used for hadronization, 
$D_{H/\tilde{g}}(z)=\delta(z-1)$.   This
assumption results in the hardest $x_{F}$ distribution at leading twist since
the $R$-hadron carries all of the gluino's momentum.  Other fragmentation
functions would soften these distributions. Note that for any
fragmentation function to factorize, it must be independent of the initial
state ({\it i.e.}\ it only depends on $z_{3}$ and not $x_{a}$).  
Thus, regardless of the fragmentation function used, all $R$-hadrons
will be decoupled from the initial state to leading twist.   

The partonic cross sections for gluino production in Eq.~(\ref{dsdxflt}) 
are \cite{DEQ}
\begin{eqnarray}
\frac{d\hat{\sigma}_{gg\rightarrow \tilde{g}\tilde{g}}}{d\hat{t}} & = &
\frac{9\pi\alpha_{s}^{2}}{4\hat{s}^{2}}
\left[ \frac{2(m_{\tilde{g}}^{2}-\hat{t})(\hat{u} - 
m_{\tilde{g}}^{2})}{\hat{s}^{2}} 
+ \frac{m_{\tilde{g}}^{2}(\hat{s} - 4m_{\tilde{g}}^{2})}{(m_{\tilde{g}}^{2} -
\hat{t})(\hat{u}-m_{\tilde{g}}^{2})} \right. \nonumber \\
& & \mbox{} \left. + \frac{(m_{\tilde{g}}^{2}-\hat{t})(\hat{u} - 
m_{\tilde{g}}^{2}) - 2m_{\tilde{g}}^{2}(m_{\tilde{g}}^{2} + 
\hat{t})}{(m_{\tilde{g}}^{2}-\hat{t})^{2}}
+ \frac{(m_{\tilde{g}}^{2}-\hat{t})(\hat{u}-m_{\tilde{g}}^{2})
+ m_{\tilde{g}}^{2}(\hat{u}-\hat{t})}{\hat{s}(m_{\tilde{g}}^{2}-\hat{t})}
\right. \\ \nonumber
& & \mbox{} \left. + \frac{(m_{\tilde{g}}^{2}-\hat{u})(\hat{t} - 
m_{\tilde{g}}^{2}) - 2m_{\tilde{g}}^{2}(m_{\tilde{g}}^{2} + 
\hat{u})}{(m_{\tilde{g}}^{2}-\hat{u})^{2}}
+ \frac{(m_{\tilde{g}}^{2}-\hat{u})(\hat{t}-m_{\tilde{g}}^{2})
+ m_{\tilde{g}}^{2}(\hat{t}-\hat{u})}{\hat{s}(m_{\tilde{g}}^{2}-\hat{u})}
\right]  \label{ggchannel} \\
\frac{d\hat{\sigma}_{q\bar{q}\rightarrow \tilde{g}\tilde{g}}}{d\hat{t}} & = &
\frac{\pi\alpha_{s}^{2}}{\hat{s}^{2}} \left[
\frac{8}{3}
\frac{(m_{\tilde{g}}^{2}-\hat{t})^{2}
+(\hat{u}-m_{\tilde{g}}^{2})^{2}
+2m_{\tilde{g}}^{2}\hat{s}}{\hat{s}^{2}} +\frac{32}{27}
\frac{(m_{\tilde{g}}^{2}-\hat{t})^{2}}{(m_{\tilde{q}}^{2}-\hat{t})^{2}}
+\frac{32}{27}
\frac{(\hat{u}-m_{\tilde{g}}^{2})^{2}}{(\hat{u}-m_{\tilde{q}}^{2})^{2}}
\right. \\ \nonumber
& & \mbox{} + \left. \frac{8}{3}
\frac{(m_{\tilde{g}}^{2}-\hat{t})^{2}+m_{\tilde{g}}^{2}\hat{s}}{\hat{s}(\hat{t}
-m_{\tilde{q}}^{2})} +\frac{8}{27}
\frac{m_{\tilde{g}}^{2}\hat{s}}{(m_{\tilde{q}}^{2}-\hat{t})
(\hat{u}-m_{\tilde{q}}^{2})}
+\frac{8}{3}
\frac{(\hat{u}-m_{\tilde{g}}^{2})^{2}+m_{\tilde{g}}^{2}\hat{s}}{\hat{s}(\hat{u}
-m_{\tilde{q}}^{2})} \right] \label{qqchannel} \\
\frac{d\hat{\sigma}_{gq\rightarrow \tilde{q}\tilde{g}}}{d\hat{t}} & = & 
\frac{\pi\alpha_{s}^{2}}{\hat{s}^{2}} \left[ \frac{4}{9}
\frac{m_{\tilde{g}}^{2}-\hat{t}}{\hat{s}}
+\frac{(m_{\tilde{g}}^{2}-\hat{t})\hat{s}+2m_{\tilde{g}}^{2}
(m_{\tilde{q}}^{2}-\hat{t})}{(m_{\tilde{g}}^{2}-\hat{t})^{2}} 
+\frac{4}{9} \frac{(\hat{u}-m_{\tilde{g}}^{2})(\hat{u}
+m_{\tilde{q}}^{2})}{(\hat{u}-m_{\tilde{q}}^{2})^{2}} \right. \\ \nonumber
&  & \mbox{} - \left. \frac{(\hat{s}-m_{\tilde{q}}^{2}
+m_{\tilde{g}}^{2})(m_{\tilde{q}}^{2}-\hat{t})
-m_{\tilde{g}}^{2}\hat{s}}{\hat{s}(m_{\tilde{g}}^{2}-\hat{t})} 
+ \frac{1}{18} \frac{\hat{s}(\hat{u}+m_{\tilde{g}}^{2})
+2(m_{\tilde{q}}^{2}-m_{\tilde{g}}^{2})
(m_{\tilde{g}}^{2}-\hat{u})}{\hat{s}(\hat{u}-m_{\tilde{q}}^{2})} \right. \\
\nonumber
& & \mbox{} + \left. 
\frac{1}{2}
\frac{(m_{\tilde{q}}^{2}-\hat{t})(2\hat{u}+m_{\tilde{g}}^{2} + 
\hat{t})}{2(\hat{t}-m_{\tilde{g}}^{2})(\hat{u}-m_{\tilde{q}}^{2})}
+ \frac{1}{2}
\frac{(m_{\tilde{g}}^{2}-\hat{t})(\hat{s}+2\hat{t} - 
2m_{\tilde{q}}^{2})}{2(\hat{t}-m_{\tilde{g}}^{2})(\hat{u}-m_{\tilde{q}}^{2})}
\right. \\ \nonumber
& & \mbox{} + \left. \frac{1}{2}
\frac{(\hat{u}-m_{\tilde{g}}^{2})(\hat{t}+m_{\tilde{g}}^{2}
+2m_{\tilde{q}}^{2})}{2(\hat{t}-m_{\tilde{g}}^{2})(\hat{u}-m_{\tilde{q}}^{2})}
\right] \, \, . \label{gqchannel}
\end{eqnarray}

\begin{figure}[p]
\phantom{space}\vfill
\setlength{\epsfxsize=0.95\textwidth}
\setlength{\epsfysize=0.5\textheight}
\centerline{\epsffile{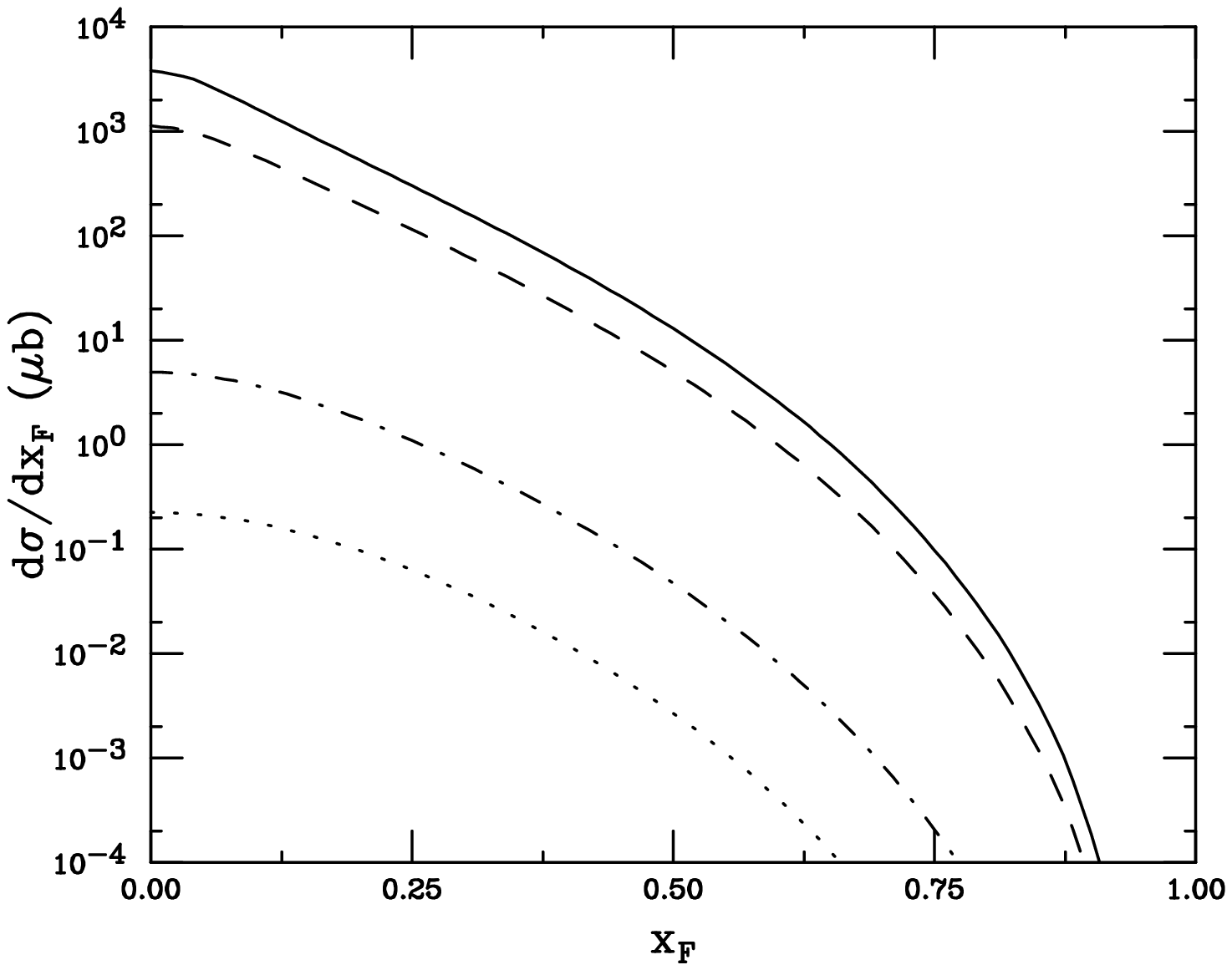}}
\vskip .75in
\caption[]{$800$ GeV QCD $pp$ gluino production for several gluino masses.  
The curves are $m_{\tilde{g}}=1.2$ GeV (solid), $1.5$ GeV (dashed), $3.5$ GeV
(dot-dashed), and $5.0$ GeV (dotted).}
\phantom{space}\vfill
\label{fu1}
\end{figure}

We calculate leading twist pQCD gluino 
distributions for $800$ GeV $pp$ interactions.  Figure~\ref{fu1} shows the
gluino distributions using the MRS D-' parton distributions in the proton
\cite{mrsdm} with $m_{\tilde{g}}=1.2, 1.5, 3.5$, and 
$5.0$ GeV and 
$m_{\tilde{q}}=100$ GeV.  The characteristic falloff at large $x_{F}$ is 
similar to heavy quark production.  Choosing a larger squark mass would
only marginally decrease the total cross section because the $gq$ channel is
suppressed by the large squark mass.
The gluino production cross section is a strong function of mass.  The cross
section is largest for $m_{\tilde{g}}=1.2$ GeV and decreases by a factor of 3
for $m_{\tilde{g}}=1.5$ GeV.  There is then a drop of 250 to the
$m_{\tilde{g}}=3.5$ GeV gluino cross section and another factor of 20 between
the $3.5$ and $5$ GeV cross sections.  Additionally, the falloff of the cross
section with $x_{F}$ becomes steeper as $m_{\tilde{g}}$ is increased.

Charm hadroproduction phenomenology has taught us that higher 
twist contributions
can become comparable to leading twist in certain parts of phase space,
introducing correlations between the initial and final states.
These effects will be addressed in the next section.

\section*{Intrinsic Contribution to Higher Twist}

In deep inelastic scattering, higher twist terms in the OPE are
suppressed by a factor of $1/Q^{2n}$.  These terms are essentially irrelevant 
when $Q^{2}$ is large.  Analogously, in hadroproduction, a similar suppression 
of $1/M^2$ typically  renders higher-twist effects unimportant except in 
regions where pQCD is seemingly inapplicable ({\it i.e.}\ 
where $M^{2}$ is small).  
However, it has been shown that in the simultaneous 
$M^{2}\rightarrow \infty$ and $x\rightarrow 1$ limit with $M^{2} (1-x)$ fixed, 
a new hard scale emerges where higher twist contributions to the cross
section become comparable to leading twist \cite{BHMT_xf,Tang,Hoy_xf}.
In the case of heavy quark production, this new scale can be associated
with either the resolution of the transverse size of the intrinsic heavy 
quark pair or with the transverse resolution of any ``pre-coalesced'' hadrons 
inside
the parent hadron.  The heavy quark fluctuations can carry a large fraction 
of the projectile's forward
momentum since the constituents of the bound state move with the same 
velocity.  
The Fock state may be broken up by an interaction with soft gluons in the
target, producing a leading hadron containing a heavy parton.

The bound state wave function for a state containing higher-twist 
contributions 
can be obtained from the 
Bethe-Saltpeter formalism evaluated at equal ``time''
on the light cone \cite{StanIC,LSIC}:
\begin{eqnarray}
(M^{2}-\Sigma_{i=1}^{n}\frac{\hat{m}_{i}^{2}}{x_{i}})
\Psi(x_{i},k_{T_i})=\int_{0}^{1} [dy] \int
\frac{[d^2l_{T}]}{16 \pi^{2}} 
\tilde{K}(x_{i},k_{T_i};y_{i},l_{T_i};M^{2})\Psi(y_{i},l_{T_i})
\label{bsform}
\end{eqnarray}
where $M$ is the mass of the projectile hadron.  The transverse mass of an 
individual parton is defined by $\hat{m_{i}}^{2}=k_{T_{i}}^2+m_{i}^2$, where
$k_{T_{i}}$ is the
transverse momentum of the $i^{\rm{th}}$ parton in the $n$-particle Fock state,
$|q_{1},...,q_{i},...,q_{n}\rangle$. 
The momentum fraction of the $i^{\rm{th}}$ parton in the
Fock state
is $x_{i}$, $[dy]=\Pi_{i=1}^{n}dy_{i}\delta(1-\Sigma_{i=1}^{n}y_{i})$ is a
longitudinal momentum conserving metric and
$[d^2l_T]=\Pi_{i=1}^n d^2l_{T_i}\delta^2(\Sigma_{i=1}^n \vec l_{T_i})$.  
The interaction kernel is $\tilde{K}$.

The simplest way to create final state hadron distributions
from a specific Fock state wave function is now described.
The vertex function on the right hand side of Eq.~(\ref{bsform}) 
is assumed to be slowly varying with
momentum.  The operator on the left hand side of the equation is then 
evaluated at the 
average transverse momentum of each parton, $\langle k_{T_{i}}^{2}\rangle$, 
with
the constraint $\Sigma_{i}^{n}\vec{k}_{T_{i}}=0$.  With these assumptions, the 
transverse mass of each
parton is fixed and the vertex function becomes constant.
The probability distribution is then proportional to the square of the
wave function which is now inversely proportional to the off-shell
parameter $\Delta=M^{2}-\Sigma_{i=1}^{n}\langle\hat{m}_{i}^{2}\rangle/x_{i}$ 
where $\langle\hat{m}^{2}_{i}\rangle$ is the average transverse mass squared of
the $i^{\rm{th}}$ parton.
After longitudinal momentum conservation is specified by
$\delta(1-\Sigma_{i=1}^{n}x_{i})$, the probability distribution becomes
\begin{eqnarray}
\frac{d^{n}P_{n}(x_{1},...,x_{n})}{\Pi_{i=1}^{n}dx_{n}} = N_{n}
\delta(1- \Sigma_{i=1}^{n}x_{i}) \Delta^{-2}
\label{dprobdx}
\end{eqnarray}
where $N_{n}$ is the normalization constant for an $n$-particle distribution.
The probability distributions as a function of $x$ for any final state hadron
can be generated by integrating Eq.~(\ref{dprobdx}) including final state 
coalescence constraints.  

The characteristic shape of the longitudinal momentum distribution of the 
final state hadron can now be obtained up to an overall normalization constant.
The important feature
of this model is that final state particles are not ``produced'' in a
collision, as such, but are rather ``intrinsic'' to the projectile's Fock state
and are liberated after a soft interaction with the target. 
This intrinsic source of final state particles acts as a perturbation to the 
dominant parton fusion mechanism.  However, unlike parton fusion, it 
incorporates flavor correlations between the initial and final states.  This 
mechanism will dominate the total cross section in the limit 
$x_{F}\rightarrow 1$ since $x_F \sim x$ when the final-state hadron evolves
directly from the projectile wave function.

In this paper, we assume that the model developed for heavy quark
hadroproduction at higher twist can be applied to gluino production in the
proton wave function.  Final-state $R$-hadron production from I$\tilde {\rm G}$
states is described in the
remainder of this section along with its relationship to IC production.
The characteristic shapes of the intrinsic distributions 
in the proton were generated for the gluino alone and for the $R^{+}(uud\tilde
g)$  and $S^{0}(uds \tilde g)$, and the $R^{0} (g \tilde g)$.  
In all cases, the ``minimal Fock state'' was used to generate the final state
coalescence.  This emphasizes the most leading final states.

The gluino can fragment into a $R$-hadron, just as in pQCD production.  In this
uncorrelated case \cite{oops}, the hadron $x_{F}$ distribution is
\begin{eqnarray}
\frac{dP_{i\tilde{g}}^{kF}}{dx_{H}}=N_{k}
\int\Pi_{j=1}^{k}dx_{j}dz\delta(1-\sum_{i=1}^{k}x_{i})
\frac{D_{H/\tilde{g}}(z)}{z}\delta(x_{H}-zx_{\tilde{g}})\Delta^{-2} \, \, ,
\label{dpdxfrag}
\end{eqnarray}
where $k$ indicates the order of the Fock state containing the intrinsic
gluinos ({\it i.e.}\ the $x_{\tilde g}$'s are included among the $x_i$). 
Gluinos are produced in pairs because other
supersymmetric vertices involving squarks and photinos are highly 
suppressed due to their much greater masses.  The minimal proton Fock state
with a gluino pair then has five particles, $|uud \tilde g \tilde g \rangle$.
Fragmentation of other, higher, Fock states will have a smaller production
probability and produce gluinos with lower average momentum.  For consistency
with the coalescence production described below, we include fragmentation of
six and seven particle Fock states with $R^0$ and $S^0$ production
respectively. 

$R$-hadron production by coalescence is specific to each hadron.  The intrinsic
gluino Fock states are fragile and can easily collapse into a new hadronic
state through a soft interaction with the target, as is the case for IC
states.  The coalescence function is assumed to be a delta function. The
momentum fraction of the of the final state hadron is the sum of the momentum
fractions of the of the $R$-hadron valence
constituents from the proton wave function.  The three $R$-hadrons we consider
are all calculated from only the minimal Fock state required for their
production by coalescence.  Thus, only the most leading configuration is used.
As in the fragmentation case in Eq.~(\ref{dpdxfrag}), including higher Fock
components does not significantly increase the total rate because the other
Fock state probabilities are smaller and also does 
not enhance the yield at large
$x_F$ because the average $x_F$ of coalescence is reduced relative to that from
the minimal Fock state.

The five-particle Fock state $|uud\tilde{g}\tilde{g}\rangle$ produces the most
leading $R$-hadron, the  $R^{+}$, because the $R^{+}$ is generated from four of
the five constituents of the Fock state.
\begin{eqnarray}
\frac{dP_{i\tilde{g}}^{5C}}{dx_{R^{+}}} =
N_{5}P^5_C \int\Pi_{j=1}^{5}dx_{j}\delta(1-\sum_{i=1}^{5}x_{i})
\delta(x_{R^{+}}-x_{u}-x_{u}-x_{d}-x_{\tilde{g}})\Delta^{-2}.
\label{dpdxrplus}
\end{eqnarray}
Here, $P^5_C$ is a factor incorporating the coalescence probability
given the five-constituent Fock state.
Note that in this case, the $R^+$ $x_F$ distribution is 
proportional to the gluino
distribution in Eq.~(\ref{dpdxfrag}), obtained by setting 
$D_{H/{\tilde g}}(z)=\delta(1-z)$,
with $k=5$ evaluated at $1-x_F$.

The $R^{0}$ is generated from a six-particle Fock state,
$|uudg\tilde{g}\tilde{g}\rangle$.  Unlike the gluinos, single gluons can be
included in the higher-twist Fock state since one gluon can couple to two 
quarks in the Fock state.  The six-particle state is the
most leading state for $R^0$ production.  
The coalescence of $R^{0}$ hadrons is described by
\begin{eqnarray}
\frac{dP_{i\tilde{g}}^{6C}}{dx_{R^{0}}} =
N_{6} P_C^6\int\Pi_{j=1}^{6}dx_{j}\delta(1-\sum_{i=1}^{6}x_{i})
\delta(x_{R_{0}}-x_{g}-x_{\tilde{g}})\Delta^{-2}.
\label{dpdxrzero}
\end{eqnarray}

The last $R$-hadron we consider is the $S^{0}$ which, since it contains an $s$
quark, must be produced from a seven-particle Fock state,
$|uuds\bar{s}\tilde{g}\tilde{g}\rangle$.  The $S^0$ will have a
harder $x_{F}$ distribution than the $R^0$ even though the average momentum 
fraction of each constituent in the seven-particle state is smaller than 
those of the six-particle state.  This harder $x_{F}$ distribution is 
due to the greater number of $S^0$ constituents, four, rather than the two 
$R^{0}$ constituents.  In this case,
\begin{eqnarray}
\frac{dP_{i\tilde{g}}^{7C}}{dx_{S^{0}}} =
N_{7} P_C^7\int\Pi_{j=1}^{7}dx_{j}\delta(1-\sum_{i=1}^{7}x_{i})
\delta(x_{S_{0}}-x_{u}-x_{d}-x_{s}-x_{\tilde{g}})\Delta^{-2}.
\label{dpdxszero}
\end{eqnarray}

In what follows, the coalescence probabilities $P_C^{5,6,7}$ appearing
in Eqs.~(\ref{dpdxrplus}), (\ref{dpdxrzero}),(\ref{dpdxszero}) 
are taken to be unity.
That is, it is assumed that the gluinos will always coalesce.

Figure~\ref{ig1} shows (using arbitrary normalization)
the characteristic $x$ dependence of the
probability distributions in Eqs.~(\ref{dpdxfrag})-(\ref{dpdxszero})
with $m_{\tilde{g}}=1.2$ GeV.   The single gluino distribution is calculated
using $k=5$ and $D_{H/{\tilde g}}(z)=\delta(1-z)$ in Eq.~(\ref{dpdxfrag}).  
$R$-hadrons produced by uncorrelated
fragmentation have the softest $x_F$ distributions,
$\langle x_{\tilde{g}}\rangle=0.24$ when $k=5$.  Contributions from 
progressively higher single gluino Fock states have smaller relative
probabilities, as we discuss below, and a decreased $\langle x_{\tilde g}
\rangle$, which would 
eventually build up a gluino sea in the proton.  The distributions from
coalescence are all forward of the single gluino distribution.
As expected, since the $R^{+}$ takes all three of the proton valence quarks, 
it is the most leading $R$-hadron with $\langle x_{R^{+}}\rangle=0.76$.  The
distributions for the other final state particles, the $S_{0}$ and the $R_{0}$,
are softer with $\langle x_{S^{0}}\rangle=0.56$ and $\langle
x_{R^{0}}\rangle=0.35$ respectively.  

We have shown the results with the lowest  gluino mass we consider.  Increasing
the mass increases the average $x_F$ of the gluino distribution of uncorrelated
fragmentation, Eq.~(\ref{dpdxfrag}), but leaves the average $x_F$ of the $R$
hadrons unchanged in the mass range we consider.

The intrinsic gluino production cross section for $R$-hadrons, from an
$n$-particle Fock state is written by analogy with the IC cross section
\begin{eqnarray}
\sigma^{n}_{i\tilde{g}}(pp)  =  G_C P^{n}_{i\tilde{g}}
\alpha_{s}^{4}(m_{\tilde{g}\tilde{g}}) \sigma^{\rm in}_{pp}
\frac{\mu^{2}}{4\hat{m}^{2}_{\tilde{g}}} \, \, , \label{sigig}
\end{eqnarray}
where $G_C$ is a color factor.
The inelastic $pp$ cross section is $\sim 35$ mb at 800 GeV. 
The ratio $\mu^{2}/4\hat{m}_{\tilde{g}}^{2}$
sets the scale at which the higher and leading twist contributions are
comparable.  We use $\mu^{2}\sim 0.2$ GeV$^{2}$, consistent with
attributing the diffractive fraction of the total $J/\psi$ production cross
section to IC \cite{BVIC1,BVIC2,GutV}.  
There is a factor of $\alpha_s^4$ because the intrinsic state
couples to two of the projectile valence quarks.  The higher-twist contribution
then contains two more powers of $\alpha_s$ than the leading-twist
contribution. This factor
is included in the cross section rather than in the 
probability distributions as done previously \cite{BVIC2,GutV} to more
explicitly show the effect of this dependence on the cross section when the
mass of the intrinsic state is changed.

Since the intrinsic charm cross section is \cite{BVIC2}
\begin{eqnarray} 
\sigma^{n}_{ic}(pp) = P^{n}_{ic}\alpha_{s}^{4}(m_{c \overline c})
\sigma^{\rm in}_{pp}
\frac{\mu^{2}}{4\hat{m}^{2}_c} \, \, , \label{sigic}
\end{eqnarray}
the two cross sections are related by
\begin{eqnarray}
\frac{\sigma^{n}_{i\tilde{g}}(pp)}{\sigma^{n}_{ic}(pp)} = 
\frac{G_CP^{n}_{i\tilde{g}}}{P^{n}_{ic}}
\frac{\hat m^2_c}{\hat{m}^{2}_{\tilde{g}}}
\frac{\alpha_{s}^4(m_{\tilde{g}\tilde{g}})}{\alpha_{s}^4(m_{c\bar{c}})}
\label{sigigicrat}
\end{eqnarray}
The relative color factor between intrinsic gluinos
and intrinsic charm, represented by $G_C$, may enhance 
the I$\tilde{\rm{G}}$ contribution over that of IC because of the color 
octet nature of the gluino. However, in this work, to isolate
mass effects, we assume the color factors for I$\tilde{\rm G}$ 
are the same as IC, setting $G_C=1$.  Changing $G_C$ would
effectively scale the cross section ratio in Eq.~(\ref{sigigicrat}) by a
constant factor.  The overall effect of changing $G_C$ is small relative
to the leading-twist cross section unless $G_C$ is very large.
The intrinsic charm mass is used as the scale from which to
approximately evolve the intrinsic gluino cross section as previously done
for intrinsic beauty \cite{BVIC2}.  Note that when $G_C = 1$, if
$\hat{m}_{\tilde g} = \hat{m}_c$, the I$\tilde{\rm G}$ and IC cross sections
are the same.  The I$\tilde{\rm G}$ cross sections are
normalized by scaling $P_{i \tilde g}$ in proportion to $P_{ic}$, as described
below.

A limit of $P_{ic}^5 = 0.31$\% was placed on the intrinsic charm probability
in the five-particle state $|uudc \overline c \rangle$ by charm
structure function data \cite{EMC_1,HM,Har}.  
The higher Fock state probabilities were
obtained from an estimate of double $J/\Psi$ production \cite{NA3psipsi}, 
resulting in $P^7_{icc}\sim 4.4\% P^5_{ic}$ \cite{BVIC5}.  
Mass scaling was used to
obtain the mixed intrinsic charm probabilities,
$P^7_{iqc}\sim (\hat{m}_{c}/\hat{m}_{q})^{2} P^7_{icc}$ \cite{BVIC1}.
To obtain the $n$-particle gluino Fock state probabilities, 
$P^{n}_{i\tilde{g}}$, 
we assume that the same relationships hold 
for the gluino states.  The five-particle
gluino state then scales as 
\begin{eqnarray} 
P^5_{i\tilde{g}} = \frac{\hat{m}_{c}^2}{\hat{m}_{\tilde{g}}^2} P^5_{ic} \, \, .
\label{pigtopic}
\end{eqnarray} 
Assuming
$P^7_{i\tilde{g}\tilde{g}}= 4.4\%P^7_{i\tilde{g}}$, 
the seven-particle Fock state probabilities are
\begin{eqnarray}
P^7_{iq\tilde{g}} =  \frac{\hat{m}_{c}^2}{\hat{m}_{q}^2}
P^7_{i\tilde{g}\tilde{g}} \, \, .
\label{piqgtopigg}
\end{eqnarray}   
Thus, if $\hat{m}_{\tilde g} = \hat{m}_c$, $P^5_{ic}
= P^5_{i \tilde g}$ and $P^7_{icc} = P^7_{i \tilde g \tilde g}$. For
simplicity, the probability for the mixed gluon-gluino proton six-particle 
Fock state was set equal to the seven-particle mixed probability with 
$\hat{m}_{g}=\hat{m}_{q}$.  The effective transverse masses used were 
$\hat{m}_{q}=\hat{m}_{g}=0.45$ GeV, $\hat{m}_{s}=0.71$
GeV, and $\hat{m}_{c}=1.8$ GeV.  The transverse mass of the gluino,
$\hat{m}_{\tilde{g}}$, is fixed to the values of $m_{\tilde{g}}$ used in the
leading twist calculation.

\section*{Composite Model Predictions}

In this section, we calculate the total $x_F$ distribution of final-state
$R$-hadrons including both leading- and higher-twist contributions.  The model
predictions for $R^+$, $R^0$ and $S^0$ production on proton and nuclear targets
are then given at 800 GeV.

The final state $d\sigma/dx_{F}$ distribution
is the sum of the leading twist pQCD distribution and the higher twist
intrinsic contributions.  Since many experiments use a nuclear target, the
characteristic $A$ dependence of each contribution is included,
\begin{eqnarray}
\frac{d\sigma}{dx_{F}}=A \frac{d\sigma_{lt}}{dx_{F}}+A^{\beta}
\frac{d\sigma_{i\tilde{g}}}{dx_{F}} \, \, .
\label{dsigsum}
\end{eqnarray}
The first term is the leading twist term whereas the second term is the higher
twist I$\tilde{\rm{G}}$ contribution.
Leading twist necessarily involves single parton interactions between the
target and the projectile and thus cannot account for collective nuclear
effects.  Thus, the leading twist cross section scales linearly with the number
of nucleons in the target modulo nuclear shadowing effects.
The nuclear dependence of $J/\psi$ production in $pA$ interactions shows 
that if the nuclear
dependence is parameterized by $A^{\alpha}$, $\alpha \rightarrow 2/3$ as
$x_{F}\rightarrow 1$ \cite{NA3_1,psiadep}. 
The emergence of this surface effect at large
$x_{F}$ is consistent with spectators in the projectile coupling to soft gluons
from the front face of the target rather than the volume.  The NA3
collaboration extracted the $A$ dependence of $J/\psi$ production at large 
$x_F$ and obtained $\beta=0.71$ in Eq.~(\ref{dsigsum}) \cite{NA3_1}.  We use
the same value of $\beta$ for charm production since the available data on the
charm $A$ dependence \cite{Appel} leads us to expect a similar $A$
dependence for charm and $J/\psi$ production at large $x_F$.

The intrinsic gluino contribution to $R$-hadron production
includes contributions from both 
hadronization of single gluinos by uncorrelated fragmentation, 
Eq.~(\ref{dpdxfrag}), and coalescence into final-state $R$-hadrons, 
described in Eqs.~(\ref{dpdxrplus})-(\ref{dpdxszero}).  That is, 
\begin{eqnarray}
\frac{dP_{i\tilde{g}}^n}{dx_F}=\xi_{1}\frac{dP^{nF}_{i\tilde{g}}}{dx_F}
+ \xi_2 \frac{dP^{nC}_{i\tilde{g}}}{dx_F}
\label{probsum}
\end{eqnarray}
where $P^{nF}_{i\tilde g}$ and $P^{nC}_{i\tilde g}$ are the 
I$\tilde{\rm{G}}$ contributions from 
fragmentation and coalescence respectively.  The parameters
$\xi_{1}$ and $\xi_{2}$ allow 
adjustment of the relative gluino fragmentation and coalescence 
contributions.  
We used single gluino fragmentation from the same Fock state as the 
coalesced hadron.  That is, for $R^{+}$, $k=5$ in Eq.~(\ref{dpdxfrag}), 
while $k=6$ for $R^{0}$ and $k=7$ for $S^{0}$.
We fix $\xi_{1}=\xi_{2}=0.5$ for simplicity.  For a more 
realistic accounting of
all possible contributions to Eq.~(\ref{probsum}) for charm production, 
see Ref.~\cite{GutV} for relative charm hadron production probabilities in the
proton. The respective fragmentation and
coalescence probability distributions in Eq.~(\ref{probsum}) are converted to 
cross sections using Eq.~(\ref{sigig}) and added to the leading twist cross 
section as in Eq.~(\ref{dsigsum}).

We calculate $R$-hadron production at 800 GeV in $pp$, $p$Be,
and $p$Cu interactions with $m_{\tilde{g}}= \hat m_{\tilde g} = 
1.2$, $1.5$, $3.5$, and $5.0$ GeV.  
Delta function fragmentation was used
for single intrinsic gluino production by uncorrelated fragmentation and for
leading twist hadronization.  That is, we take $D_{H/\tilde{g}}(z)=\delta(1-z)$
in Eqs.~(\ref{dsdxflt}) and (\ref{dpdxfrag}).

\begin{figure}[p]\phantom{space}\vfill
\setlength{\epsfxsize=0.95\textwidth}
\setlength{\epsfysize=0.5\textheight}
\centerline{\epsffile{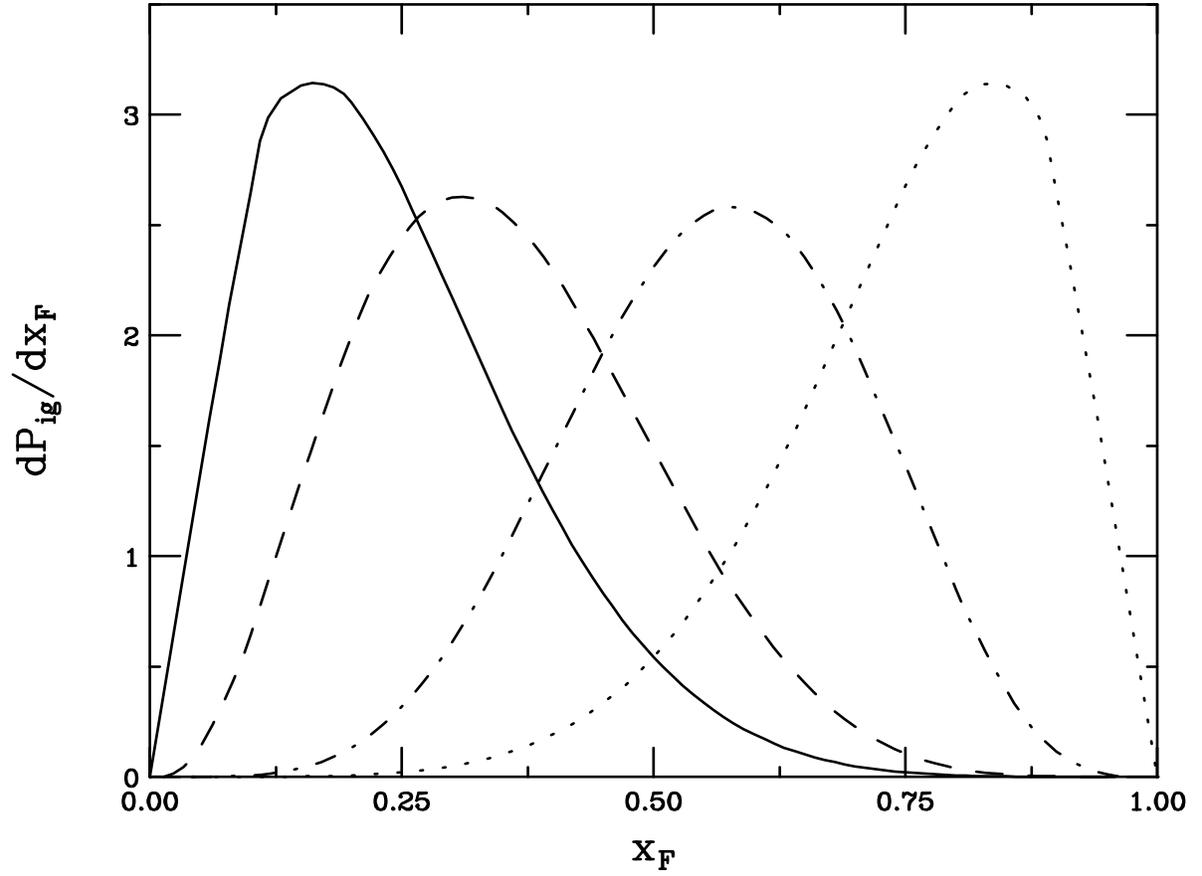}}
\vskip .75in\caption[]{The $x$ distribution of intrinsic $R$-hadrons in the proton with 
$m_{\tilde{g}}=1.2$ GeV.  The curves are $\tilde{g}$ (solid), $R^{0}$
(dashed), $S^{0}$ (dot-dashed), and $R^{+}$ (dotted).}
\phantom{space}\vfill\label{ig1}
\end{figure}

\begin{figure}[p]\phantom{space}\vfill
\setlength{\epsfxsize=0.95\textwidth}
\setlength{\epsfysize=0.5\textheight}
\centerline{\epsffile{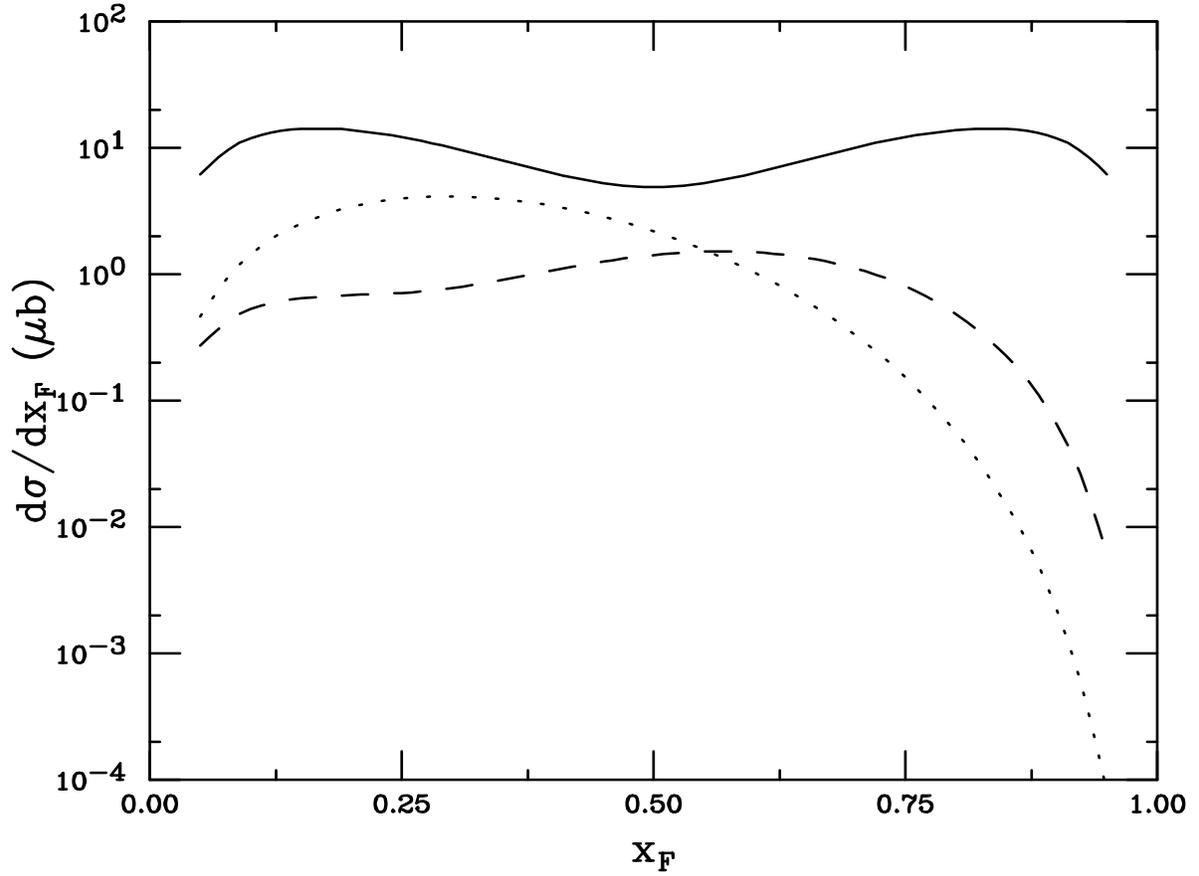}}
\vskip .75in\caption[]{Intrinsic gluino higher twist contributions to
$d\sigma_{i\tilde{g}}/x_{F}$ 
in $R$-hadron production with $m_{\tilde{g}}=1.2$ GeV.  The solid curve is
$R^{+}$, the dotted curve is $R^{0}$, and the dashed
curve is $S^{0}$.  Each distribution includes the contribution from 
independent uncorrelated fragmentation of a gluino.}
\phantom{space}\vfill\label{ig2}
\end{figure}

Figure~\ref{ig2} shows the normalized $R$-hadron $x_F$ distributions calculated
according to Eq.~(\ref{sigig}) in $pp$ interactions with $m_{\tilde g} = 1.2$
GeV.  The difference in the yields as $x_F \rightarrow 0$ is due to the
difference in probability for the five, six, and seven particle Fock states.
The $R^0$ and $S^0$ cross sections are similar at low $x_F$ because we have
assumed $P_{i g \tilde g}^6 = P_{i q \tilde g}^7$, as described in the previous
section.  However, the shapes are different at low $x_F$ because the
probability distribution for uncorrelated fragmentation has a smaller average
$\langle x_F \rangle$ when $k=7$ in Eq.~(\ref{dpdxfrag}).  The $R^+$ has the
largest cross section of the three hadrons.  Its distribution is symmetric
around $x_F = 0.5$ because the fragmentation yield and the $R^+$ yield from
coalescence are symmetric in the five particle Fock state.  The $S^0$ yield
increases near $x_F \sim 0.25$ due to the forward peak of the $S^0$ coalescence
distribution seen in Fig.~\ref{ig1}.  The yield at low $x_F$ is relatively
reduced because the fragmentation calculation with $k=7$ is narrower so that
the two peaks are effectively separated in Fig.~\ref{ig1}.  Since the
fragmentation peak for $k=6$ and the $R^0$ coalescence distribution lie close
together, they blend into a broad peak for the $R^0$ $x_F$ distribution.

\begin{figure}[p]\phantom{space}\vfill
\setlength{\epsfxsize=0.95\textwidth}
\setlength{\epsfysize=0.5\textheight}
\centerline{\epsffile{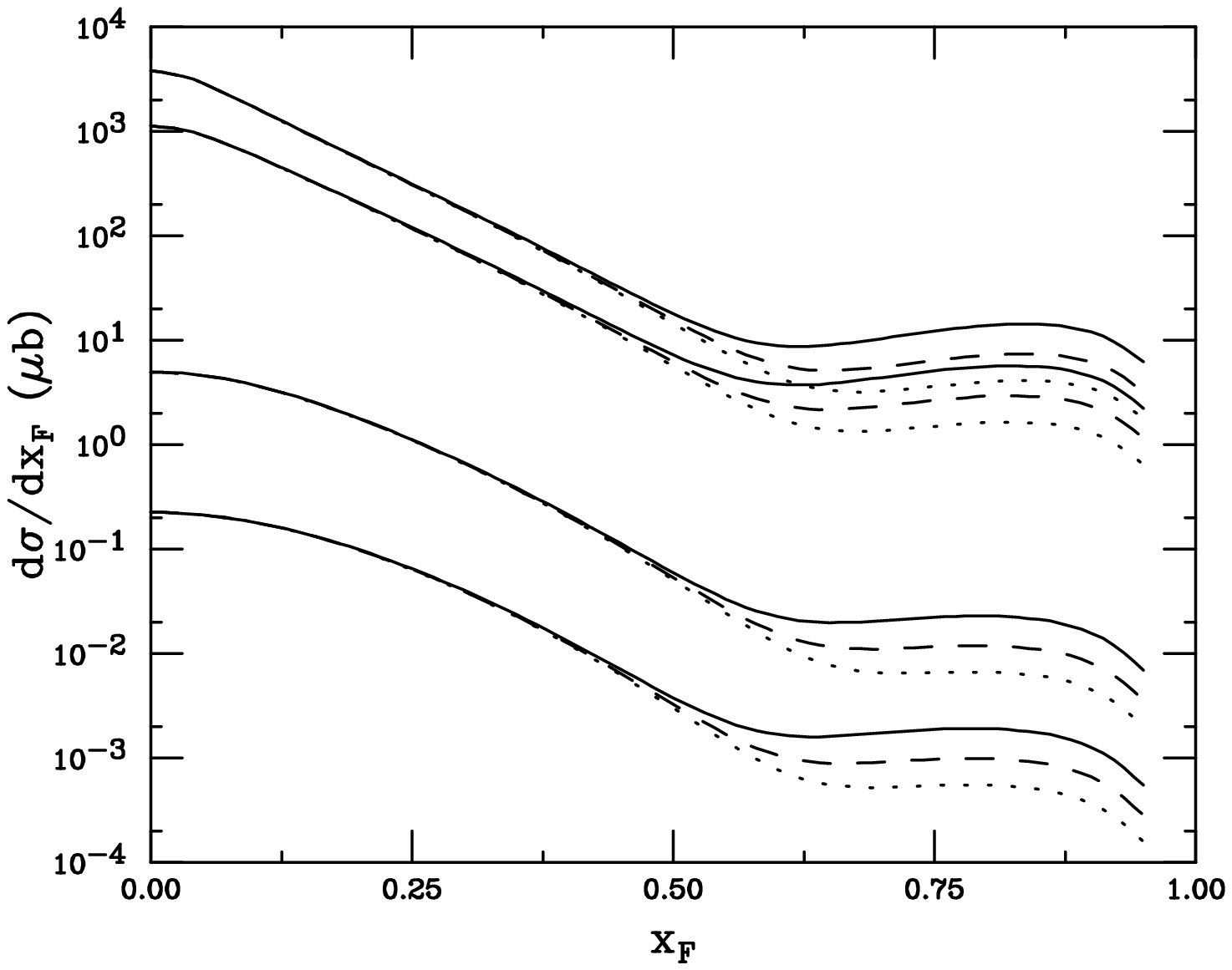}}
\vskip .75in\caption[]{$R^{+}$ $x_{F}$ distribution from $800$ GeV protons on various
targets.  Four gluino masses are chosen, $m_{\tilde{g}}=1.2$ GeV
(top), $1.5$ GeV, $3.5$ GeV, and $5.0$ GeV (bottom).  For each mass, there
is a triplet of curves representing different targets: proton (solid), Be
(dashed), and Cu (dotted).}
\phantom{space}\vfill\label{pp800Rp}
\end{figure}

\begin{figure}[p]\phantom{space}\vfill
\setlength{\epsfxsize=0.95\textwidth}
\setlength{\epsfysize=0.5\textheight}
\centerline{\epsffile{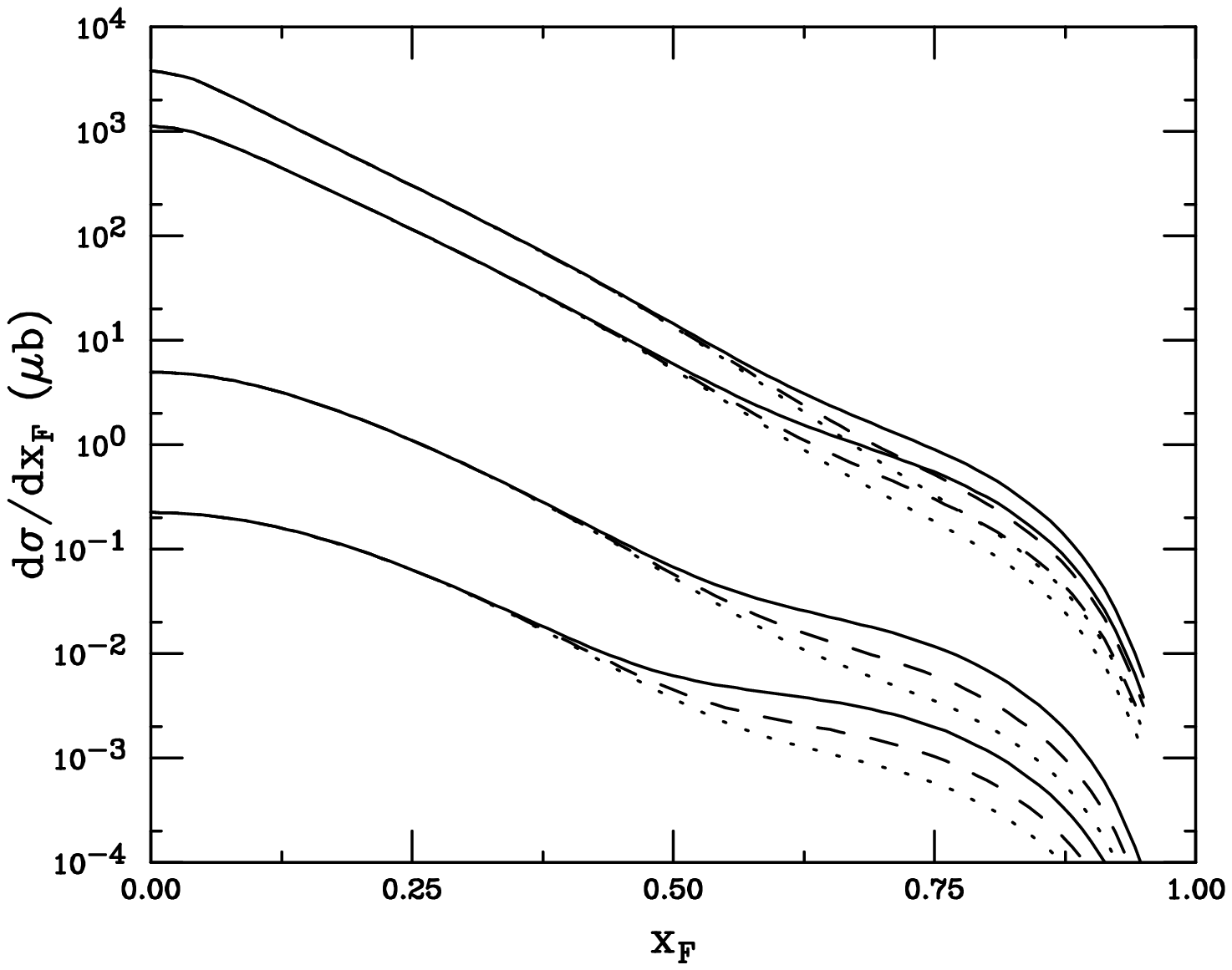}}
\vskip .75in\caption[]{$S^{0}$ $x_{F}$ distribution from $800$ GeV protons on various
targets.  Four gluino masses are chosen, $m_{\tilde{g}}=1.2$ GeV
(top), $1.5$ GeV, $3.5$ GeV, and $5.0$ GeV (bottom).  For each mass, there
is a triplet of curves representing different targets: proton (solid), Be
(dashed), and Cu (dotted).}
\phantom{space}\vfill\label{pp800S0}
\end{figure}

\begin{figure}[p]\phantom{space}\vfill
\setlength{\epsfxsize=0.95\textwidth}
\setlength{\epsfysize=0.5\textheight}
\centerline{\epsffile{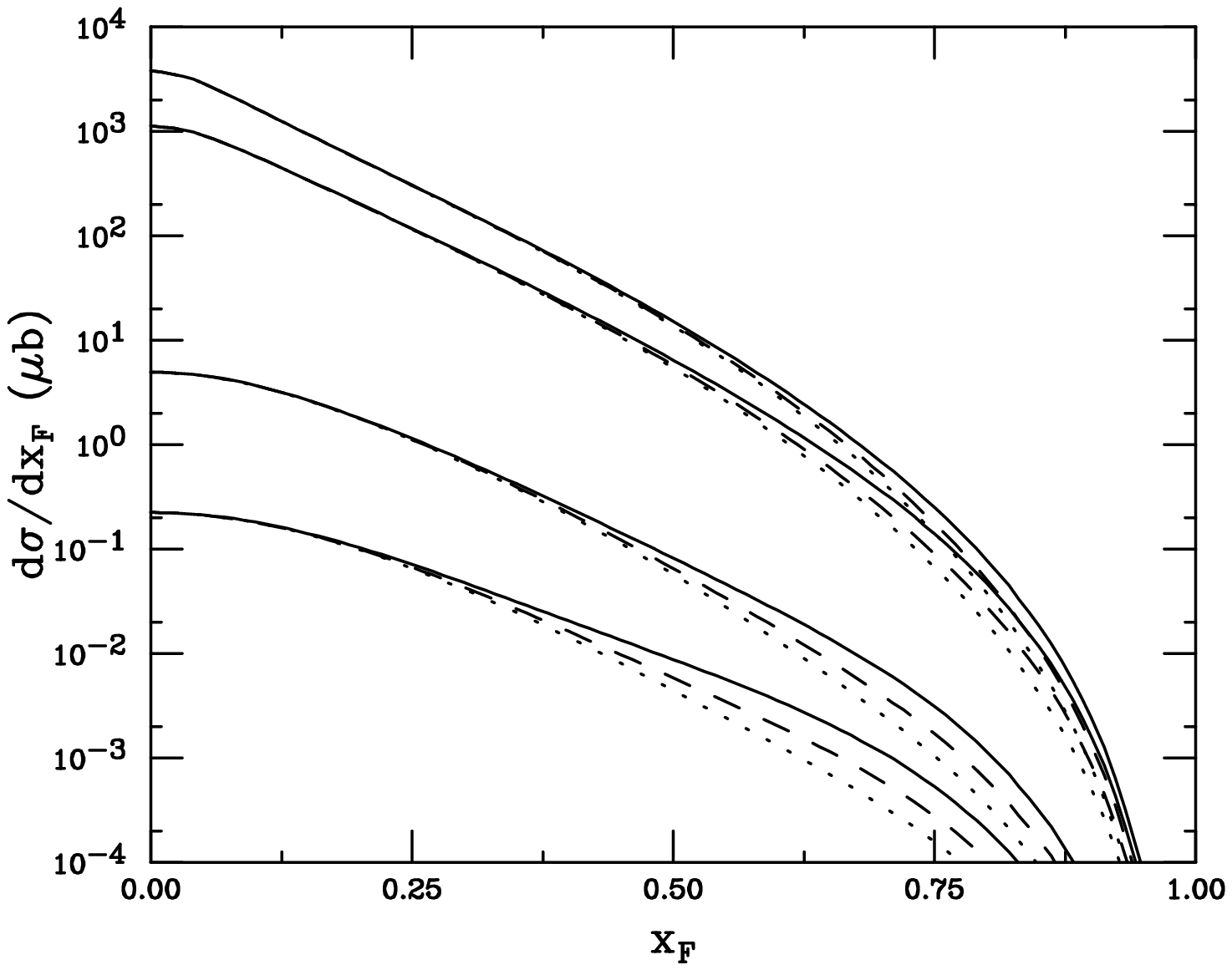}}
\vskip .75in\caption[]{$R^{0}$ $x_{F}$ distribution from $800$ GeV protons on various
targets.  Four gluino masses are chosen, $m_{\tilde{g}}=1.2$ GeV
(top), $1.5$ GeV, $3.5$ GeV, and $5.0$ GeV (bottom).  For each mass, there
is a triplet of curves representing different targets: proton (solid), Be
(dashed), and Cu (dotted).}
\phantom{space}\vfill\label{pp800R0}
\end{figure}

Figures~\ref{pp800Rp},~\ref{pp800S0}, and~\ref{pp800R0} show the predicted 
$R^{+}$, $S^{0}$, and $R^{0}$ $x_{F}$ distributions per nucleon in $pp$,
$p$Be, and $p$Cu interactions at 800 GeV calculated
according to Eq.~(\ref{dsigsum}).  Each figure
includes all four gluino masses.  As $x_{F}\rightarrow 0$ the
$x_{F}$ distributions of all targets are equal for a given
$m_{\tilde{g}}$.  This indicates the dominance of leading twist production at
low $x_{F}$, independent of the final state.  As $x_{F}\rightarrow 1$
the higher twist terms begin to contribute.  These
higher twist effects are suppressed in nuclear targets  
because of their slower relative growth as a function of $A$ compared to the
leading twist $A$ dependence.
Although larger mass gluinos are more difficult to create, the relative
contribution to the total cross section from higher-twist production in
Eq.~(\ref{dsigsum}) increases with gluino mass because of the slower decrease
of the intrinsic gluino contribution relative to the mass suppression of the
leading twist cross section. 
The greater mass suppression of the leading twist cross section also influences
the value of $x_F$ where the higher twist contribution begins to appear.
Increasing  the gluino mass leads to intrinsic gluino effects appearing at
lower $x_F$.  This effect is seen in Figs.~\ref{pp800Rp}-\ref{pp800R0}.  When
$m_{\tilde{g}}=1.2$, I$\tilde{\rm{G}}$ effects become obvious near 
$x_{F}\sim 0.5$ while I$\tilde{\rm G}$ contributions begin to appear for
$x_F \sim 0.2$ in $R^0$ production when $m_{\tilde{g}}=5.0$ GeV.

Dramatic leading effects are predicted for the $R^{+}$ which, as pointed out
above, shares three valence quarks with the proton in a minimal five-particle 
Fock state configuration.  This characteristic
``hardening'' of the $x_{F}$ distribution for $x_{F}>0.6$ should be clear 
in a successful $R^{+}$
search.  However, the leading effects are also present for the other
particles.  The $S^{0}$ is the next hardest distribution, sharing two
valence quarks with the proton while the $R^{0}$ tends to be the softest,
since no projectile valence quarks are shared.

\begin{figure}[p]\phantom{space}\vfill
\setlength{\epsfxsize=0.95\textwidth}
\setlength{\epsfysize=0.5\textheight}
\centerline{\epsffile{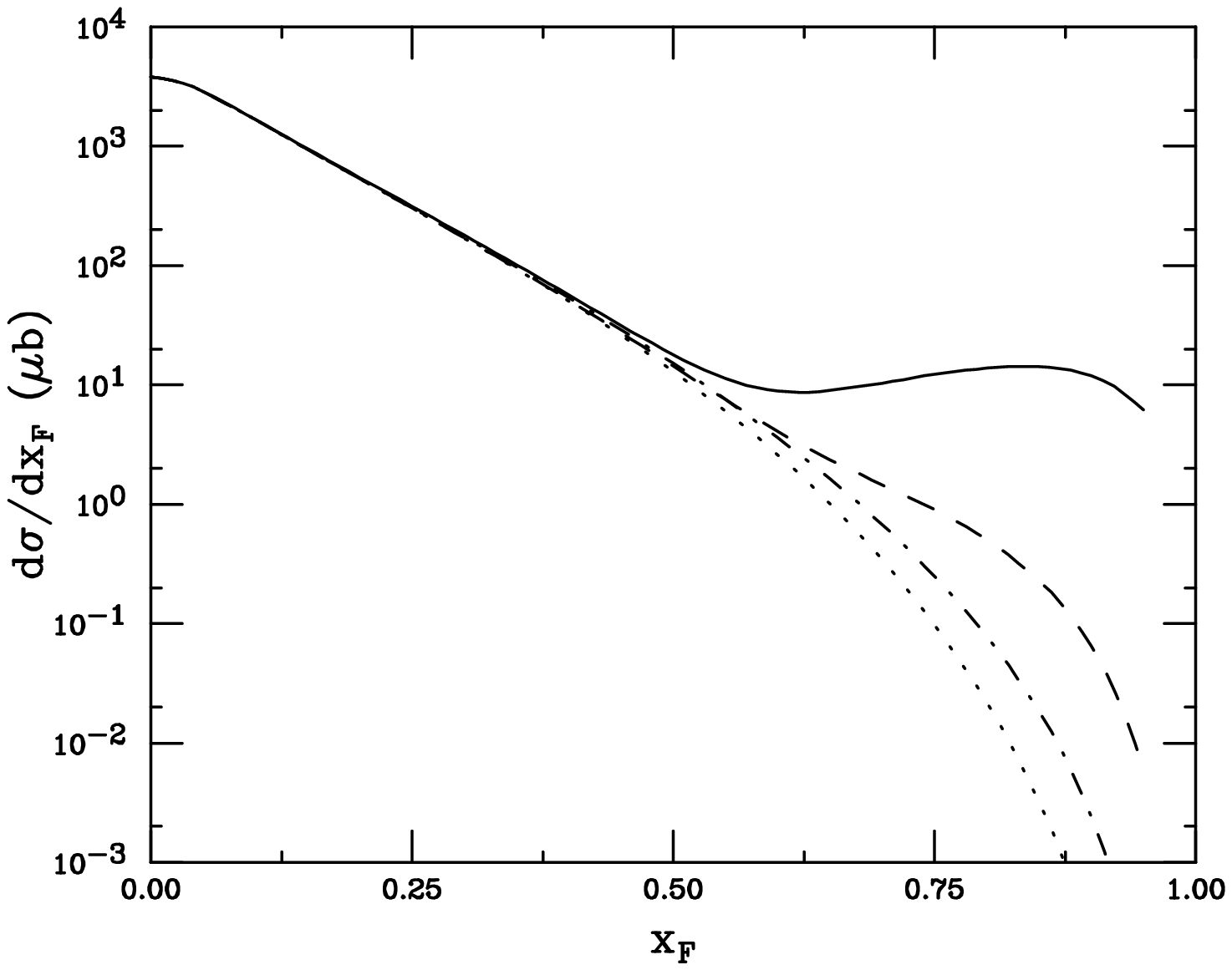}}
\vskip .75in\caption[]{Intrinsic gluino enhancement to $x_{F}$ distribution for various 
$R$-hadrons with $m_{\tilde{g}}=1.2$ GeV.  The lower curve is the fusion 
baseline for gluino production with a
delta function fragmentation.  The curves are $R^{+}$ (solid), $S^{0}$
(dashed), 
$R^{0}$ (dot-dashed), and the leading twist gluino production (dotted).}
\phantom{space}\vfill\label{pp800fi12}
\end{figure}

\begin{figure}[p]\phantom{space}\vfill
\setlength{\epsfxsize=0.95\textwidth}
\setlength{\epsfysize=0.5\textheight}
\centerline{\epsffile{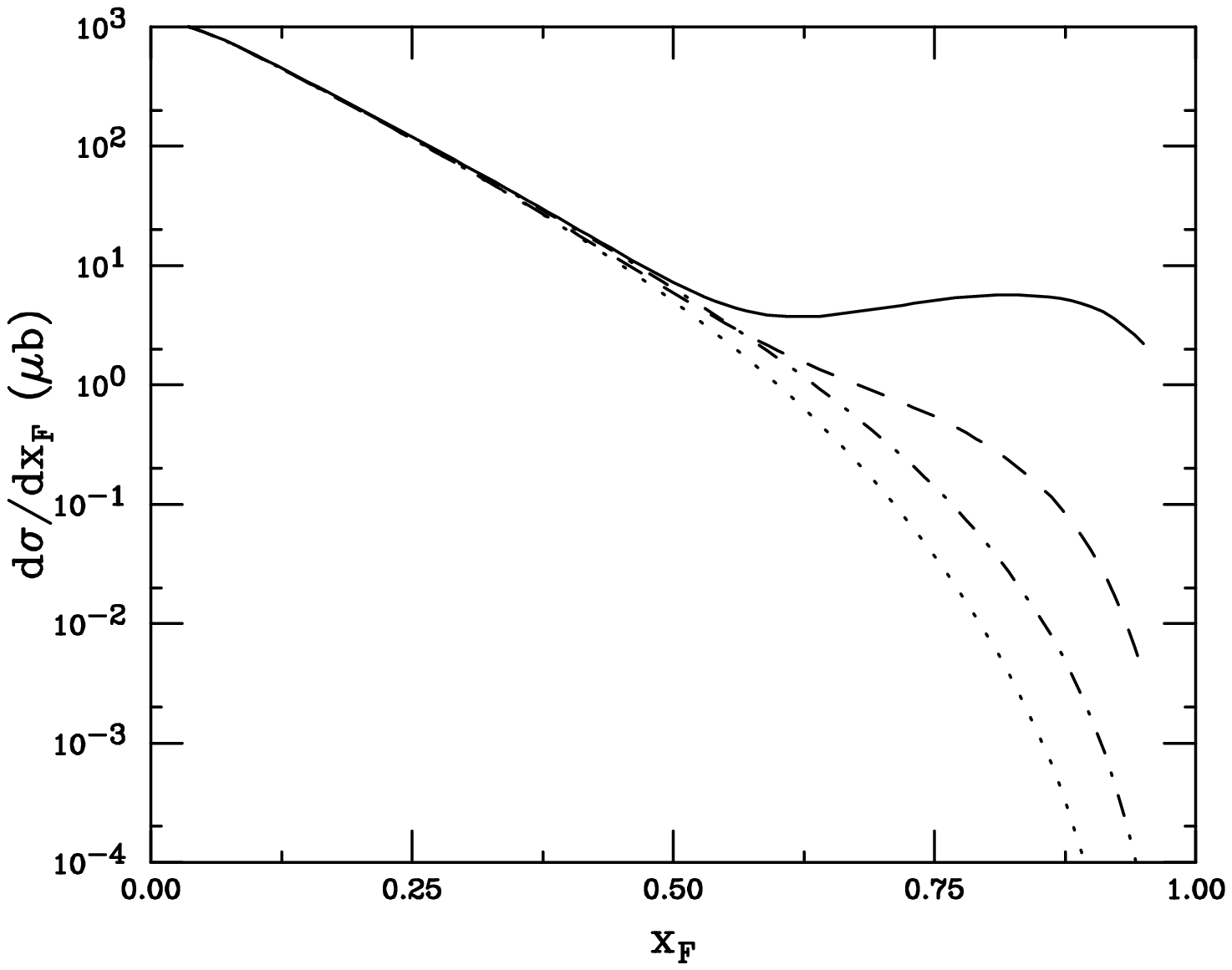}}
\vskip .75in\caption[]{Intrinsic gluino enhancement to $x_{F}$ distribution for various 
$R$-hadrons with $m_{\tilde{g}}=1.5$ GeV.  The lower curve is the fusion 
baseline for gluino production with a
delta function fragmentation.  The curves are $R^{+}$ (solid), $S^{0}$
(dashed), 
$R^{0}$ (dot-dashed), and the leading twist gluino production (dotted).}
\phantom{space}\vfill\label{pp800fi15}
\end{figure}

\begin{figure}[p]\phantom{space}\vfill
\setlength{\epsfxsize=0.95\textwidth}
\setlength{\epsfysize=0.5\textheight}
\centerline{\epsffile{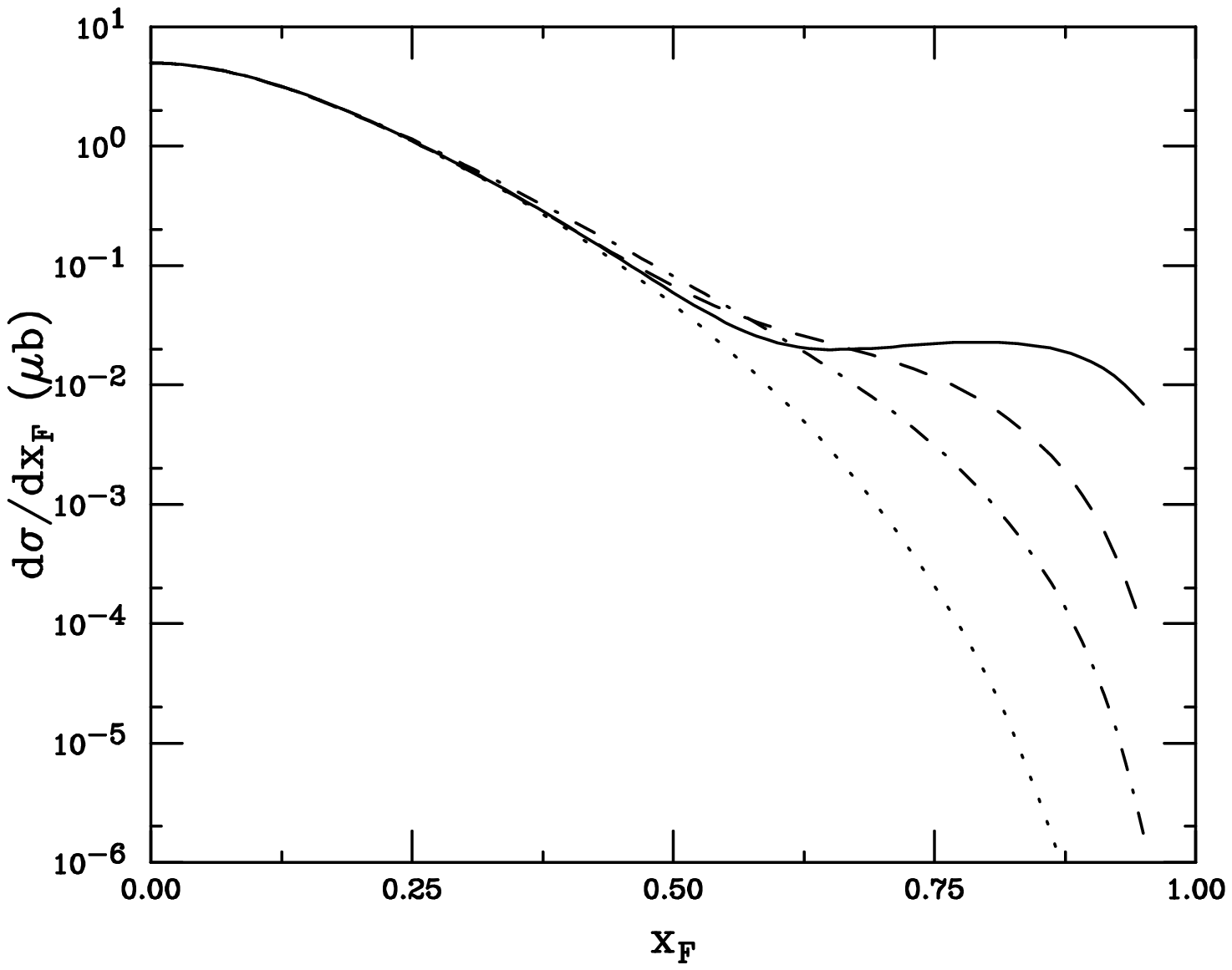}}
\vskip .75in\caption[]{Intrinsic gluino enhancement to $x_{F}$ distribution for various 
$R$-hadrons with $m_{\tilde{g}}=3.5$ GeV.  The lower curve is the fusion 
baseline for gluino production with a
delta function fragmentation.  The curves are $R^{+}$ (solid), $S^{0}$
(dashed), 
$R^{0}$ (dot-dashed), and the leading twist gluino production (dotted).}
\phantom{space}\vfill\label{pp800fi35}
\end{figure}

\begin{figure}[p]\phantom{space}\vfill
\setlength{\epsfxsize=0.95\textwidth}
\setlength{\epsfysize=0.5\textheight}
\centerline{\epsffile{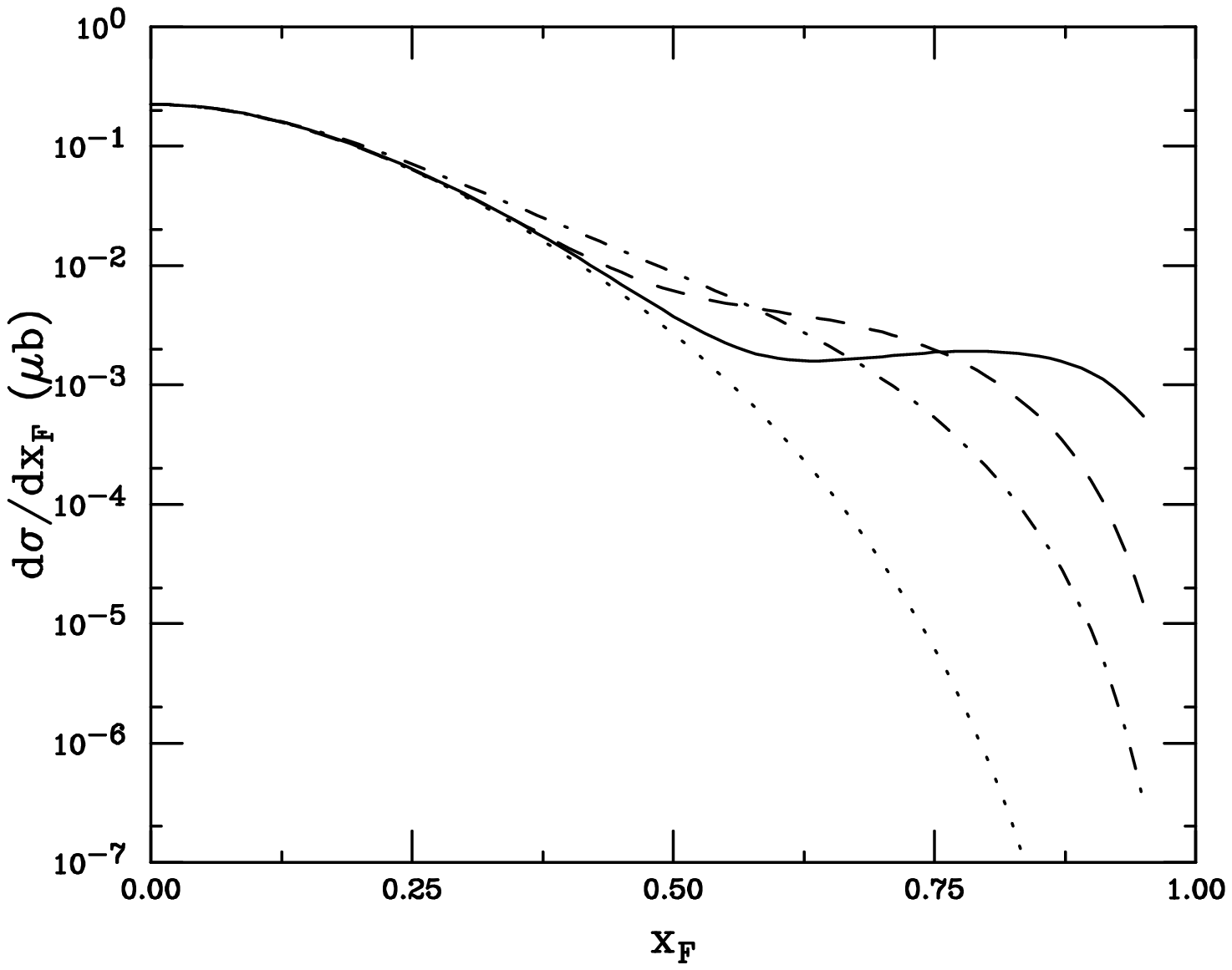}}
\vskip .75in\caption[]{Intrinsic gluino enhancement to $x_{F}$ distribution for various 
$R$-hadrons with $m_{\tilde{g}}=5.0$ GeV.  The lower curve is the fusion 
baseline for gluino production with a
delta function fragmentation.  The curves are $R^{+}$ (solid), $S^{0}$
(dashed), 
$R^{0}$ (dot-dashed), and the leading twist gluino production (dotted).}
\phantom{space}\vfill\label{pp800fi50}
\end{figure}

For a clearer comparison of the leading
effects predicted for each final state $R$-hadron,
Figs.~\ref{pp800fi12}-\ref{pp800fi50} show the 
$R^{+}$, $S^{0}$, and $R^{0}$ distributions together in $pp$
interactions with $m_{\tilde{g}}=1.2, 1.5, 3.5$, and $5.0$ GeV respectively.
The leading twist gluino distribution is also shown for comparison.  
In each case, the intrinsic
contribution begins to emerge from the leading twist calculation
between $x_{F}\sim 0.2$ and $x_{F}\sim 0.4$.  
In Fig.~\ref{pp800fi15}, with $m_{\tilde{g}}=1.5$ GeV, the
predicted $R^{+}$ enhancement at $x_{F}\sim 0.8$ is about $700$ 
times larger than the leading twist prediction.  At the same value of 
$m_{\tilde{g}}$ and $x_{F}$, the $S^{0}$ contribution is about $40$ times 
greater while the $R^{0}$ is just under $6$ times greater.
When the gluino mass is increased to 
$m_{\tilde{g}}=5.0$ GeV, shown in Fig.~\ref{pp800fi50}, the $R^{0}$ dominates 
$R$-hadron yields for $x_{F}<0.6$.  This is a consequence of the increased
$\langle x_F \rangle$ for single gluino fragmentation at the larger mass.
Although the cross sections are small at $m_{\tilde{g}}=5.0$ GeV since the
gluino mass is comparable to the bottom mass,
the predicted enhancements over the leading-twist baseline
are quite large:  $2.5\times10^{3}$ for the $R^{+}$, $1.6\times10^{3}$ for the 
$S^{0}$, and $281$ for the $R^{0}$.  The enhancements are in fact larger than
those with smaller gluino masses due to the greater mass suppression of the
leading twist cross section.

\section*{Conclusions}

The light gluino window opens the possibility of non-trivial higher 
twist gluino contributions to the proton wave function.  In analogy to charm 
hadroproduction, intrinsic gluino Fock components contribute to final state
$R$-hadron formation, enhancing gluino production over leading
twist parton fusion in the forward $x_F$ region.  

In this work, we have studied a ``maximally leading'' scenario for final state
$R$-hadrons in $pp$ and $pA$ interactions at $800$ GeV.  
Our model predicts that the contributions of higher-twist 
intrinsic states lead to  strong flavor correlations between initial and
final states for $x_{F}>0.6$.  The large intrinsic gluino
enhancements at high $x_F$ over the leading-twist predictions 
imply that this region of phase space could be especially appropriate
for $R$-hadron searches in the light gluino scenario. 
For $m_{\tilde g}$ in the $1-5\gev$ range, a mass region where substantial
evidence for the analogous intrinsic heavy quark states exists and for which
our computational techniques should be most reliable, the enhancements
are very significant (factors of several hundred to several thousand 
being common).
The magnitudes we predict for these enhancements may even be conservative since
the increased color factor associated with intrinsic gluinos compared
to intrinsic charm has been neglected.\\[2ex]

{\bf Acknowledgements} We would like to thank S.J. Brodsky for discussions.


\begin{references}
\bibitem{oops}
The factor of $1/z$ associated with the fragmentation function was left out
in Refs.~\cite{BVIC1,BVIC2,BVIC4,GutV}.  However, the factor was included in
computing the numerical results given in these papers.
\end{references}
\end{document}